# A Surface Acoustic Wave based Single Photon Shifter for Solid-state Sources

Jiaxiang Guo, Huijun Zhao, Kaili Xiong, Pingxing Chen, Yang Zhang[*], Yan Chen[*]

**Abstract:** Controlling the frequency of nonclassical light is indispensable for implementing quantum computation, communication and bridging various quantum systems. However, frequency-shift devices for solid state single-photon sources that are easy to integrate are practically absent. Here, we propose an integrated single-photon frequency shifter based on acousto-optic modulation. The device consists of two Interdigital Transducers (IDTs) for surface acoustic wave (SAW) generation and a silicon waveguide periodically placed at the nodes the SAW to increase the interaction length. The $V_\pi{*}L$ of the device is $1.2v.cm$. Under 133.2MHz driving frequency and 10 volt driving voltage, a shift up to $\pm65.7$GHz is achieved with near unity conversion efficiency. Our results demonstrate the feasibility of on-chip deterministic quantum spectral control in constructing hybrid quantum networks.

## 1. Introduction

Solid state single photon emitters including color centers, quantum dots, defects in low dimensional materials etc. have played an important role in photonic quantum information processing [1,2]. Despite significant progresses have been achieved to make single emitters optimal, the frequency mismatch among emitters poses a major challenge. The frequency mismatch lies at the fabrication methods which are inherently random. Individual emitters face different electro-magnetic environments, making them dissimilar with each other. However, most photonic applications rely on photon interference-based measurements [3]. The dissimilarity of emitters prevents its scalability. To circumvent this problem, time demultiplexing of one solid state single photon emitter is widely adopted. [4-7]. This sets a high requirement on the source brightness. Tremendous efforts are therefore devoted to improve the source brightness. Still the photon number of multiphoton experiments is substantially hampered. Furthermore, there are growing needs to interface systems of different kinds to make a hybrid system. To take full advantage of the excellent performance of solid state single photon emitters, their frequency control has been an integral part of a wide range of photonic quantum technologies. There are many ways to control the frequency of solid-state single photon emitters. At the source part, by growth optimazition[8-10], the inhomogeneous broadening was narrowed down to several nanometers. Alternatively, post growth tuning techniques including E-field tuning, magnetic field tuning as well as strain engineering are widely deployed [11]. Although it is now feasible to engineer the spectrum of individual emitters, the preparation complexity of the sources is increased, making their integration with other functional circuits remains challenging. It is desired to tune the frequency of single photon emitters after the photons are emitted, with the sources being intact.

Usually, the conversion of single photons' frequencies involves nonlinear optical processes, imposing strict requirements for momentum and energy conservation[12-21]. Recently, single photon shifters based on Electro-optical modulation (EOM)[22,23] .and optomechanical chips provide an important means to for spectrum engineering[24,25]. Among these schemes, the magnitude of frequency shift depends on the modulation frequency $v_m$ and driving amplitude $v_0$[22]. To obtain observable frequency shift, all the working modulation exceeds ~GHz, corresponding to a temporal span of tens of picoseconds. This temporal duration is much smaller than the lifetime of a photon emitted from solid states, which would inevitably result in the spectrum distortion via fast modulation.

In the paper, we propose a SAW based single photon shifter, which is tailored for photons emitted from solid state photon sources. The shifter consists of a SAW modulator and a folded silicon waveguide periodically placed at the nodes of SAW. The SAW is set to work at low modulation frequency of 133MHz to avoid spectrum distortion. The SAW introduces strain on the silicon waveguide, and changes its refractive index by photo-elastic effects. The $\pi$ voltage length product of the device is 1.2 *v.cm.*. Frequency shift of $\pm 65.7$GHz is achieved at 10V modulation amplitude.

## 2. Methods

**Device structure and principle.** Figure 1(a) shows schematic diagram of the acousto-optic frequency shifter. The two IDTs are placed onto a host substate consisting of a thin layer lithium niobate. In order to alleviate the clamping effect, the thin layer of lithium niobate beneath the IDTs has been intentionally designed to be suspended. The spacing between the two IDTs is integer multiple of their pitch dimensions (equal to the wavelength of SAW,$\lambda$) to form standing acoustic wave and enhance mechanical deformation. A silicon waveguide placed at the nodes of the SAW experiences the largest strain-induced deformation. As a result of photo-elastic effect [26,27] its refractive index is modulated accordingly with the driving signal. To avoid spectrum distortion, the light pulse duration is much shorter than the SAW modulation period, corresponding to an adiabatic process. Following the mechanical vibration, the effective refractive index changes periodically. Thus, photons in the waveguide experience different optical length changes with different drive phase $\varphi$ with respect to the photon arrival time. According to the Doppler effect, the frequency of photons will correspondingly undergo blue shift or red shift, as shown in the Figure 1(b). To improve the amplitude of frequency shift, periodically folded silicon waveguide is placed at the nodes of the SAW as is shown in Figure 1(c). The whole silicon waveguide is cladded by a silicon dioxide for better strain transfer.

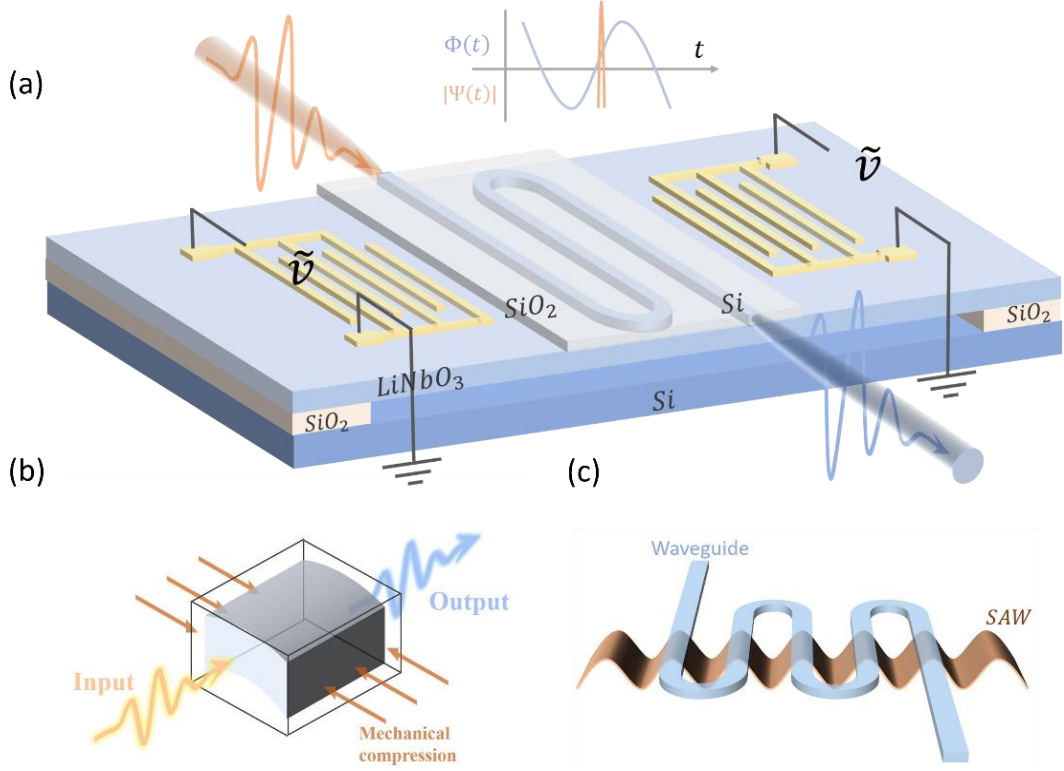

**Figure 1** (a) The structure diagram of opto-mechanical frequency shift device. The silicon waveguide experience a varying modulation. (b) Light pulses passing through a mechanical deformed waveguide undergo a frequency shift; (c) The waveguide is periodically folded to increase the modulation efficiency.

When the photon pulse duration $\tau$ is much smaller than the SAW period ($\tau \ll 1/\omega_m$), the photon number is almost unaffected. This is an intrinsic nature the adiabatic frequency shifter. In the adiabatic limit, the relationship between the photon frequency change $\delta\omega$ and the mechanical displacement $\delta x$ is given by [21]

$$\delta\omega \approx -v_g k_0 \frac{\partial n_{eff}}{\partial x} \delta x \qquad (1)$$

where $k_0$, $v_g$ and $n_{eff}$ are the photon vacuum vector, group velocity, and effective refractive index, respectively. After the establishment of standing waves, periodic fluctuations occur in the effective refractive index of the straight segments of the photonic waveguide. Consequently, distinct driving phases ($\varphi$) result in varying optical lengths of the waveguide. The displacement can be expressed as:

$$\delta x = \delta X \cos\varphi \sin\frac{\omega_m L}{2v_g} \qquad (2)$$

where $\delta X$ is the mechanical motion amplitude; $\omega_m$ is the angular frequency of SAW; $L=N \times L_s$ is the waveguide length with $N$ corresponding to the number of straight waveguides, and $L_s$ the length each straight waveguide that coincides with the SAW.

In the following, we evaluate the refractive change of silicon waveguide caused by elastic strain. The applied elastic strain field $S$ causes a change in the dielectric tensor $\varepsilon$ of silicon, which is given by[24]

$$\Delta\varepsilon_{il} = \varepsilon_{il} p_{jkmn} \varepsilon_{kl} S_{mn} \qquad (3)$$

where $p_{jkmn}$ is an element of the photo-elastic tensor. In cubic media, there are only three independent photo-elastic coefficients. For silicon $(p_{11}, p_{12}, p_{44}) = (-0.09, 0.017, -0.051)$. In our device, the photo-elasticity is mainly induced from in-plane displacement, and the change of refractive index of silicon waveguides can be written as

$$\delta n(x,y) \approx -\frac{1}{2} n_0^3 (p_{11} S_{xx} + p_{12} S_{yy}) \qquad (4)$$

where $n_0$ is the equilibrium refractive index of the material. By combining equations (1), (2), and (4), we derive the relationship between frequency shift and elastic strain.

3. Results

The IDTs used in simulation have a pitch size of 30μm, and the substrate is made of piezoelectric material lithium niobate, with a thickness of 3.5μm. For solid state light sources with a typical lifetime on the order of the nano-second, the driving frequency should be controlled at 1GHz or below. We here set the eigenfrequency of the Rayleigh surface wave generated by this IDT to be 133.2 MHz, and it can be adjusted by modifying the pitch size. The refractive index of $SiO_2$ is set to 1.445, and that of high-resistance Si is set to 3.5. To simulate the standing wave effect of SAW, Rayleigh damping is introduced.

We examine the standing wave effect of SAW within the present structure by the finite element method. Figure 2(a) and (b) depict the transient displacement and stress distribution at the eigenfrequencies, induced by sinusoidal RF signals with a 10V amplitude in the Y direction of lithium noibate. The SAW exhibits stable behavior on the thin lithium niobate layer, displaying prominent periodicity and well-defined boundaries. Subsequently, we conduct simulations to analyze the propagation characteristics of photonic waveguides at the communication band. Our photonic waveguides demonstrate remarkable transmittance over a broad spectral range of 1.45-1.65 μm. Figure 2(c) presents the transmission, absorption, and reflection spectra. The inset illustrates the field distribution of the photonic waveguide. Figure 2(d) provides an enlarged cross section of device. The silicon waveguide is placed at the nodes of the SAW. Therefore, it experiences the largest deformation or strain.

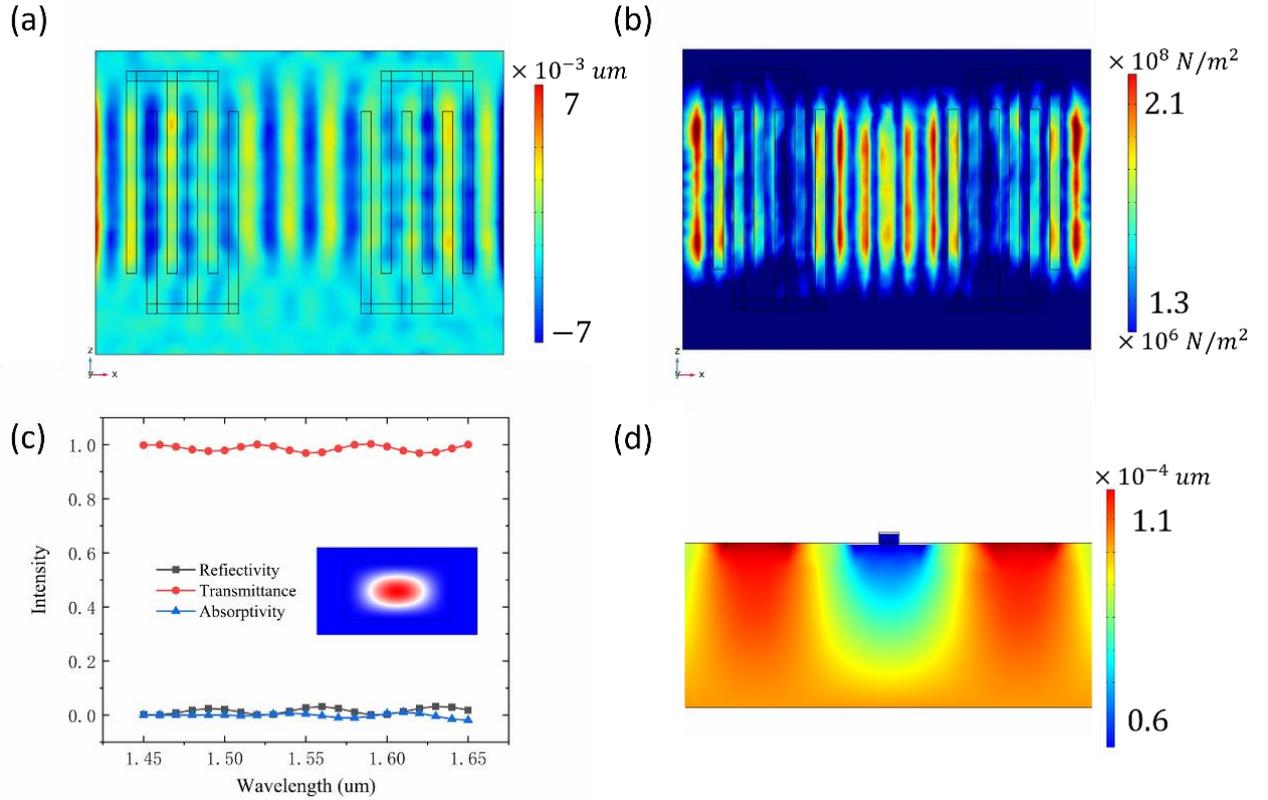

**Figure 2**. (a)The device displacement and (b)stress distribution in the Y direction when standing wave occurs; (c) Transmission, reflection and absorption spectra of photonic waveguide in the range of 1.45-1.65 μm. Inset: The electric field distribution in the waveguide; (d) A cross view of deformation distribution when the waveguide is strained.

Driven by a sinusoidal RF signal, the waveguide experiences periodic variations in its effective refractive index, leading to periodic changes in the optical path as illustrated in Figure 3(a). The mode of electric field is also plotted when $\varphi = -0.5\pi$ and $\varphi = 0.5\pi$. At these specific locations, the waveguide experiences its maximum stretching or compression. These alterations in the waveguide's physical properties cause photon pulses exhibiting redshift or blueshift. We analyze the simulated output spectrum of a Gaussian input photon under various $\varphi$ conditions. As predicted by equation (1) and (2), frequency change $\delta\omega$ follows a sinusoidal dependence on $\varphi$ for short photon duration. The maximum frequency shift and minimum shape distortion happen at $\varphi = 0$ and $\varphi = \pi$, where the slope of refractive index modulation is steepest. When $\varphi = 0$, the photon undergoes blue shift, while for $\varphi = \pi$, the photon experiences redshift. Under 133.2MHz driving frequency, 3cm waveguide lengths and 10 volt driving voltage, the shifted spectra together with the original spectra are plotted in Figure 3(b). The center wavelengths are extracted in Figure 3(c). The frequency shift undergoes a sinusoidal change with a maximum range of ± 1 nm (± 65.7 GHz). It worth to note that this shift frequency is obtained with a much low driving frequency of 133.2MHz compared with previous works.

To comprehensively explore the acousto-optic interaction modulation, we further analyze the influence of both voltage and waveguide length on the maximum frequency shift. Figure 3(d) demonstrates the frequency shift as a function of waveguide lengths, driven at 133.2MHz and 10V with a mechanical driving phase of $\pi$. As expected, the shift increases linearly with waveguide length. Figure 3(e) plots the frequency shift caused by different voltages with a fixed 30 mm waveguide length. The frequency shift exhibits a linear dependence on the driving amplitude.

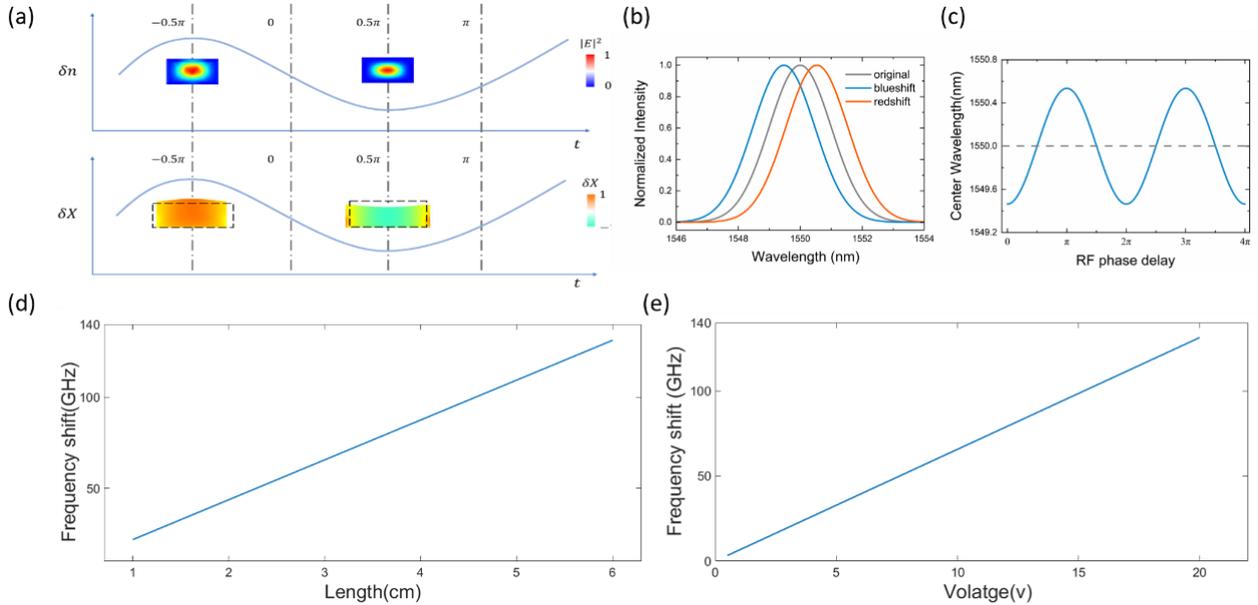

**Figure 3**. (a) The change of refractive index δn and displacement δx at different drive phases; (b) Photon frequency undergo blue/red shift at the rising/falling slopes of thesinusoidal RF drive. (c) Frequency shift as a function of RF phase delay, The modulator was driven at 133.2MHz with an amplitude of 10V. (d)-(e) The resulting frequency shift varies linearly with the amplitude and frequency of the applied voltage

Spectral control of quantum light is widely investigated through electrical driven devices. Table 1 presents a comparison of our work with select experiments from the literature. The mechanism is based on either electro-optic (EO) or acousto-optic (AO) modulation. Notably, all prior demonstrations have relied on high frequency modulators, which is not applicable for solid state quantum light sources. In fact, all these works use photons generated through nonlinear optical process. These photons have ultra short temporal duration. In our study, we aim to change the frequency of photons emitted by solid state sources. Although the observed shift is relatively modest, it is important to highlight that our driving frequency is significantly lower as exemplified by the recent study by Zhu and colleagues [24], The observed frequency shift is an order of magnitude smaller compared to their work. It is worth noting, though, that our experiment employs a driving frequency two orders of magnitude lower than theirs.

**Table 1.** Comparation of Performance in frequency Control of Quantum light sources

| Reference | Methods | Source | Wavelength (nm) | Freq. shift (GHz) | Driving Freq. (GHz) |
|---|---|---|---|---|---|
| Wright(2017)[22] | Bulk electro-optic modulation | Pulsed SPDC | 832 | $\pm 2\pi * 200$ | 40 |
| Fan(2016)[24] | Integrated opto-mechanical modulation | Pulsed SPDC | 1554 | $\pm 2\pi * 150$ | 8.3 |
| Chen(2021)[14] | Bulk electro-optic modulation | Narrow-band SPDC | 1560 | $\pm 2\pi * 15.65$ | 15.65 |
| Zhu(2022)[23] | Integrated electro-optic modulation | Pulsed SPDC | 1560 | $\pm 641$ | 27.5 |
| **This work** | Integrated acousto-optic modulation | On-chip deterministic quantum spectral | 1550 | $\pm 65.7$ | 0.1332 |

## 4. Conclusion

Based on the results presented above, we hereby propose a scheme for upscaling solid-state emitters through the utilization of acousto-optic interactions, as depicted in Figure 4. In this scheme, sources emitting light with mismatched colors are directed towards each input port. By employing different voltages on these IDTs, the output light from the corresponding port can be precisely tuned to achieve spectral consistency. The primary objective of this device is to mitigate the frequency/bandwidth differences and inhomogeneities observed among various photon sources, thereby facilitating seamless integration with spectrally mismatched quantum memories and light sources. The successful implementation of spectrum resonance through acousto-optic interaction holds significant importance for the large-scale integration of quantum systems.

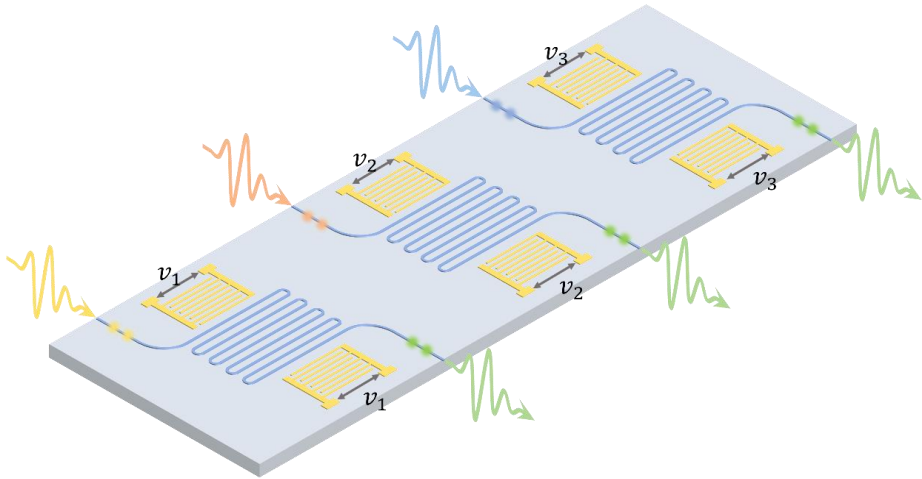

**Figure 4**. The artistic sketch of a spectrum shift device based on acousto-optic modulation. The inhomogenerous broadening of various solid state emitters are compensated via a single-photon frequency shifter.

Our approach of frequency shifter for solid state sources is generic and holds broad applicability to various substrates and photon emitters. Specifically, we propose a novel design scheme for an LNOI (Lithium Niobate-on-Insulator) optical shifter, utilizing SAW. Periodically folded low-loss photonic waveguides are placed at the nodes of the SAW where the strain change rate is maximal when driving at relatively low frequency ~133MHz. It worth to note that optical length enhancement can also be achieved through the utilization of photonic resonators, such as the race-track resonator. Nonetheless, the race-track resonator exhibits a limited bandwidth. In contrast, folded waveguides are wavelength-insensitive and offer a significantly wider bandwidth.

## Author Information

### Corresponding Authors

**Yang Zhang** − *Institute of Modern Optics, Nankai University, Tianjin Key Laboratory of Micro-Scale Optical Information Science and Technology, Tianjin, 300071 China;* Email: yangzhang@nankai.edu.cn

**Yan Chen** − *Institute for Quantum Science and Technology, College of Science, National University of Defense Technology, Changsha 410073, China; College of Advanced Interdisciplinary Studies, National University of Defense Technology, Changsha 410073, China;* Email: chenyanxyz@outlook.com

### Authors

**Jiaxiang Guo** − *Institute of Modern Optics, Nankai University, Tianjin Key Laboratory of Micro-Scale Optical Information Science and Technology, Tianjin, 300071 China; College of Advanced Interdisciplinary Studies, National University of Defense Technology, Changsha 410073, China*

**Huijun Zhao** − *Institute of Modern Optics, Nankai University, Tianjin Key Laboratory of Micro-scale Optical Information Science and Technology, Tianjin 300350, China*

**Kaili Xiong** − *Institute for Quantum Science and Technology, College of Science, National University of Defense Technology, Changsha 410073, China ; Hunan Key Laboratory of Quantum Information Mechanism and Technology, National University of Defense Technology, Changsha 410073, Hunan, China*

**Pingxing Chen** − *Institute for Quantum Science and Technology, College of Science, National University of Defense Technology, Changsha 410073, China; Hunan Key Laboratory of Quantum Information Mechanism and Technology, National University of Defense Technology, Changsha 410073, Hunan, China; Hefei National Laboratory, Hefei 230088, China*

All authors have given approval to the final version of the manuscript.

**Jiaxing guo and Huijun Zhao contribute equally.**

### Notes

The authors declare no competing financial interest.

**Acknowledgment**

We are grateful for financial support from the National Natural Science Foundation of China (12374476), the Natural Science Foundation of Hunan Province (2021JJ20051), the science and technology innovation Program of Hunan Province (2021RC3084), and the research program of national university of defense technology (ZK21-01, 22-ZZCX-067).